\documentclass[9pt, a4, margin=0.5in]{article}
\usepackage[utf8]{inputenc}
\usepackage{array}
\usepackage{graphicx}
\usepackage{subcaption}
\usepackage[english]{babel}
\usepackage{xcolor}
\usepackage{authblk}
\usepackage{geometry}
\usepackage{amsfonts}

\title{\textbf{Using spatial-temporal ensembles of convolutional neural networks for lumen segmentation in ureteroscopy}
\thanks{This work was supported by the ATLAS project. This project has received funding from the European Union’s Horizon 2020 research and innovation programme under the Marie Skłodowska-Curie grant agreement No 813782.
}}

\author[1,2]{Jorge F. Lazo}
\author[3]{Aldo Marzullo}
\author[4,5]{Sara Moccia}
\author[6]{Michele Catellani}
\author[2]{Benoit Rosa}
\author[2]{Michel de Mathelin}
\author[1]{Elena De Momi}

\affil[1]{\small{\textit{DEIB}, Politecnico di Milano, Milan, Italy}}
\affil[2]{\small{\textit{ICube, UMR 7357}, CNRS-Université de Strasbourg, Strasbourg, France}}
\affil[3]{\small{\textit{Department of Mathematics and Computer Science}, University of Calabria, Rende, Italy}}
\affil[4]{\small{\textit{The BioRobotics Institute}, Scuola Superiore Sant’Anna, Pisa, Italy}}
\affil[5]{\small{\textit{Department of Excellence in Robotics and AI}, Scuola Superiore Sant’Anna, Pisa}}
\affil[6]{\small{\textit{Istituto Europeo di Oncologia (IRCCS)}, Milan, Italy}}
\date{}

\begin{document}
\maketitle

\begin{abstract}

\mbox{}\\
\textbf{Purpose:}
Ureteroscopy is an efficient endoscopic minimally invasive technique for the diagnosis and treatment of upper tract urothelial carcinoma (UTUC). 
During ureteroscopy, the automatic segmentation of the hollow lumen is of primary importance, since it indicates the path that the endoscope should follow. 
In order to obtain an accurate segmentation of the hollow lumen, this paper presents an automatic method based on Convolutional Neural Networks (CNNs).

\noindent %
\textbf{Methods:}
The proposed method is based on an ensemble of 4 parallel CNNs to simultaneously process single and multi-frame information.
Of these, two architectures are taken as core-models, namely U-Net based in residual blocks($m_1$) and Mask-RCNN($m_2$), which are fed with single still-frames $I(t)$.
The other two models ($M_1$, $M_2$) are modifications of the former ones consisting on the addition of a stage which makes use of 3D Convolutions to process temporal information. $M_1$, $M_2$ are fed with triplets of frames ($I(t-1)$, $I(t)$, $I(t+1)$) to produce the segmentation for $I(t)$.


\noindent %
\textbf{Results:}
The proposed method was evaluated using a custom dataset of 11 videos (2,673 frames) which were collected and manually annotated from 6 patients. 
We obtain a Dice similarity coefficient of 0.80, outperforming previous state-of-the-art methods.

\noindent %
\textbf{Conclusion:}
The obtained results show that spatial-temporal information can be effectively exploited by the ensemble model to improve hollow lumen segmentation in ureteroscopic images. 
The method is effective also in presence of poor visibility, occasional bleeding, or specular reflections. 



\end{abstract}

\section{Introduction}

Upper Tract Urothelial Cancer (UTUC) is a sub-type of urothelial cancer which arises in the renal pelvis and the ureter. The disease, has an estimated number of 3,970 patients affected in 2020~\cite{Siegel2020} in the United States.
Flexible Ureteroscopy (URS) is nowadays the gold standard for UTUC diagnosis and conservative treatment. URS is used to inspect the tissue in the urinary system, determine the presence and size of tumour~\cite{cosentino2013upper} as well as for biopsy of suspicious lesions~\cite{rojas2013low}. The procedure is carried out under the visual guidance of an endoscopic camera~\cite{wason2020ureteroscopy}.

Navigation and diagnosis through the urinary tract are highly dependent upon the operator expertise~\cite{delaRosette_2006}. 
For this reason, the current development of methods in Computer Assisted Interventions (CAI) intends to support surgeons by providing them with relevant information during the procedure~\cite{munzer2018content}. 
Additionally, within the endeavours of developing new tools for robotic ureteroscopy, a navigation system which relies on image information from the endoscopic camera is also needed~\cite{borghesan2020atlas}.

\begin{figure}[tbp]
    \begin{center}
     \begin{subfigure}[b]{0.99\textwidth}
     \centering
         \includegraphics[width = 0.65\textwidth]{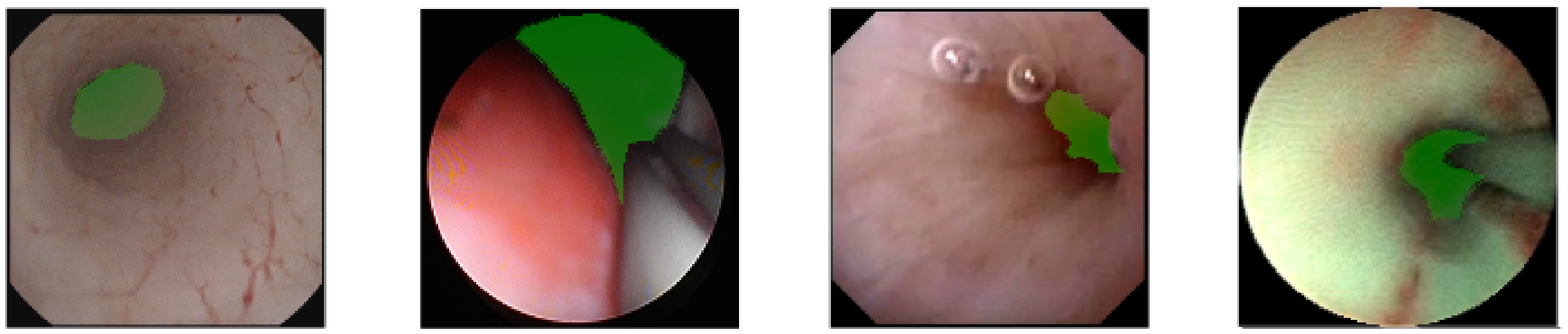} 
         \caption{\footnotesize{Variations in the shape of the lumen, and the hues of the surrounding tissue.}}
     \end{subfigure}
    \begin{subfigure}[b]{0.3\textwidth}
     \centering
         \includegraphics[width = 0.99\textwidth]{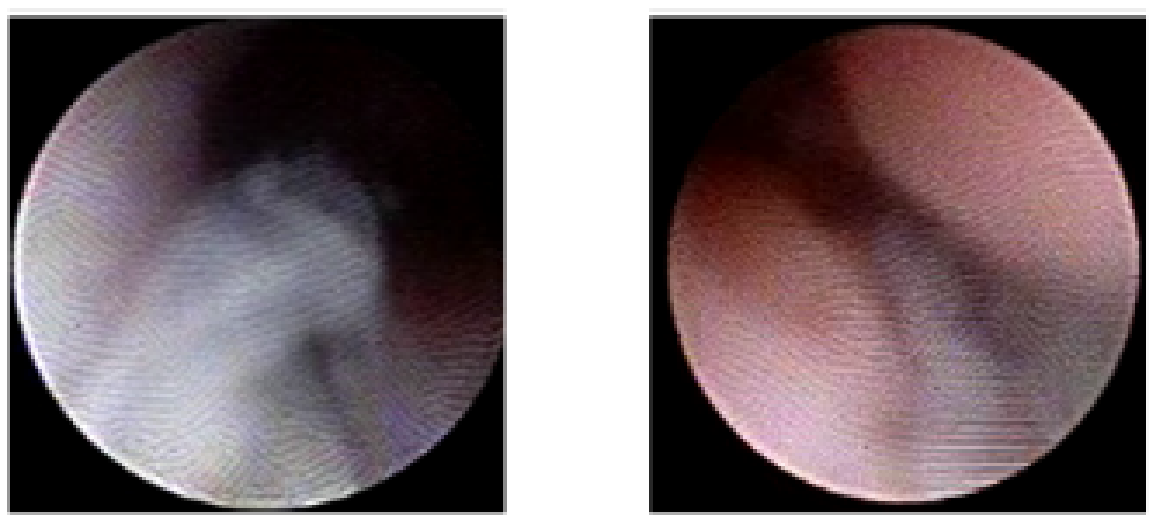} 
         \caption{\footnotesize{Noise}}
     \end{subfigure}
     \hspace{0.2cm}
    \begin{subfigure}[b]{0.3\textwidth}
     \centering
         \includegraphics[width = 0.99\textwidth]{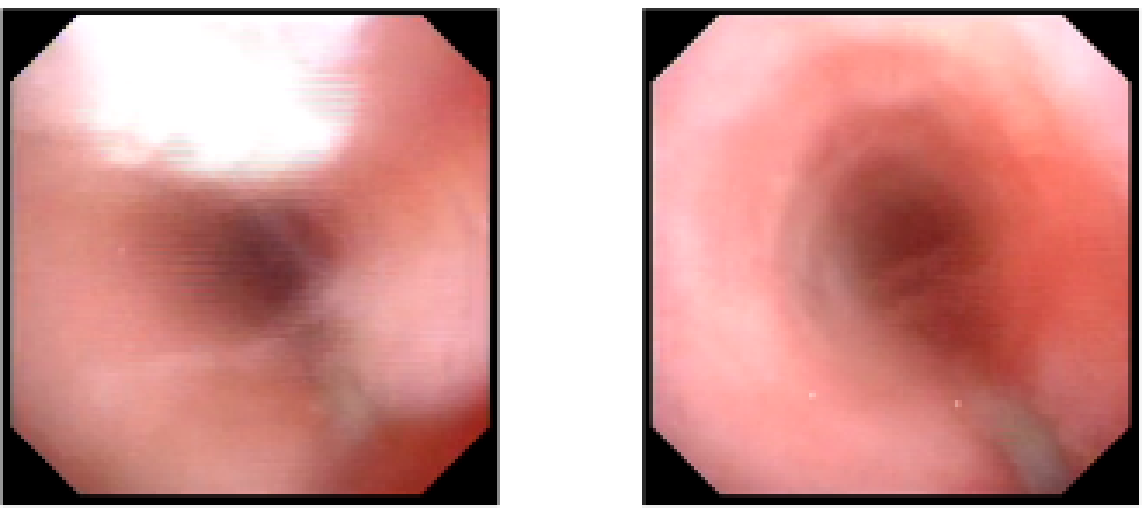} 
         \caption{\footnotesize{Blood occlusion}}
     \end{subfigure}
     
     \begin{subfigure}[b]{0.3\textwidth}
     \centering
         \includegraphics[width = 0.99\textwidth]{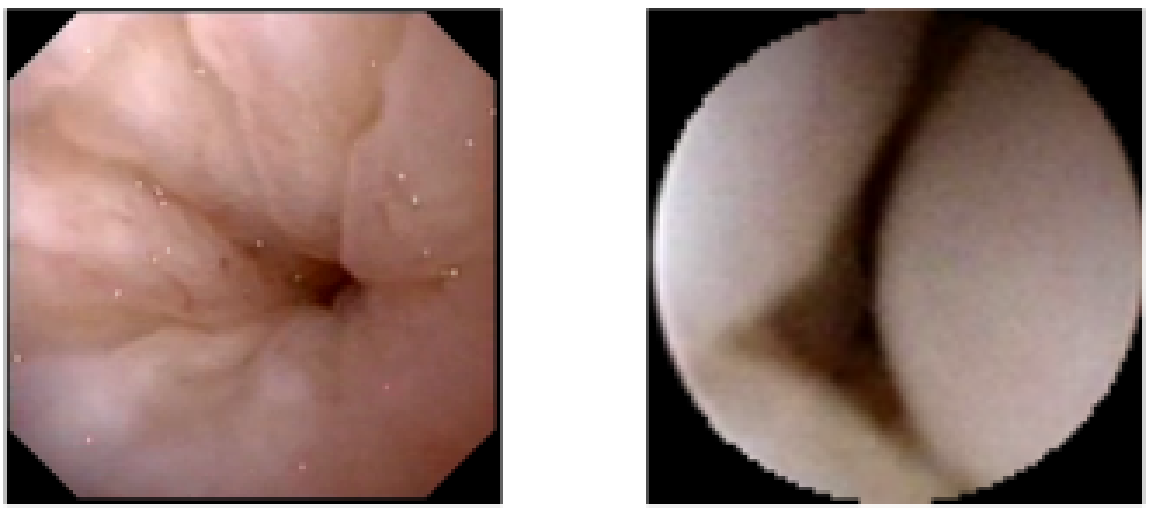} 
         \caption{\footnotesize{Lumen narrowing}}
     \end{subfigure}
     \hspace{0.2cm}
    \begin{subfigure}[b]{0.3\textwidth}
     \centering
         \includegraphics[width = 0.99\textwidth]{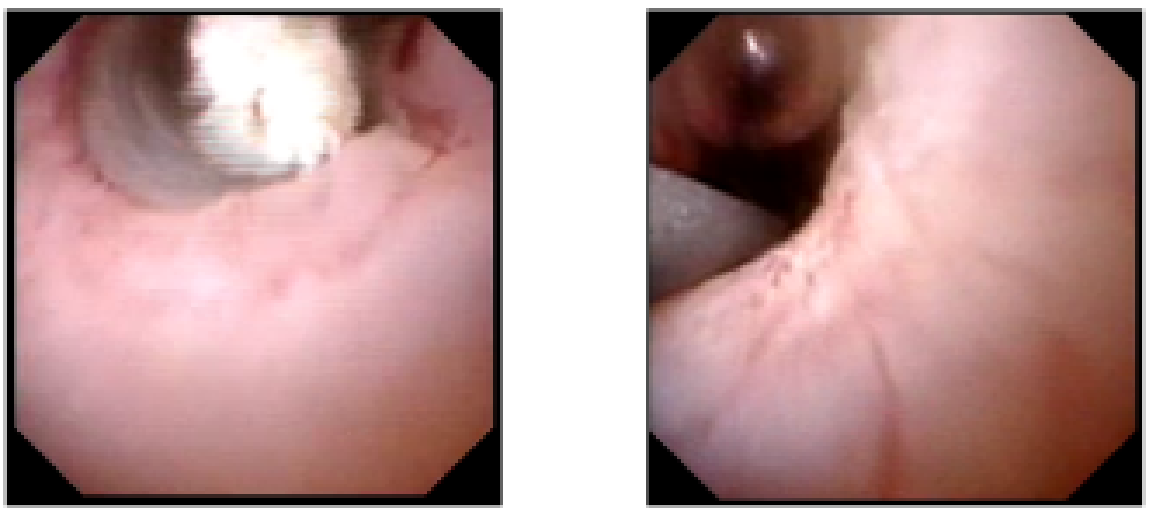} 
         \caption{\footnotesize{Debris and bubbles}}
     \end{subfigure}
    \caption{\footnotesize{Sample images in our dataset showing:  (a) the hue variability of the surrounding tissue as well as the shape and location of the lumen (the hollow lumen is highlighted in green to show clearly the variety of shapes in which it could appear). (b)-(e) Samples of artifacts (the lumen was not highlighted to have a clear view of the image artifacts).}}
    \label{fig:sample_dataset}
    \end{center}
\end{figure}

In this study, we focus on the segmentation of the ureter's lumen. 
In ureter-endoscopic images, the lumen appears most likely as a tunnel or hole in the images with its center being the region with the lowest illuminance inside the Field of View (FOV). 
Lumen segmentation presents some particular challenges such as the difficulty of defining the concrete boundary of it, the narrowing of the ureter around the ureteropelvic junction  \cite{wason2020ureteroscopy}, and the appearance of image artifacts such as blur, occlusions due to the appearance of floating debris or bleeding    . Some examples of these, present in our data, are shown in Fig.~\ref{fig:sample_dataset}.
%


In the CAI domain, Deep Learning (DL)-based methods, represent the state-of-the-art for many image processing tasks, including segmentation.  
In~\cite{vazquez2017benchmark} an 8-layer Fully Convolutional Network (FCN) is presented for semantic segmentation of colonoscopy images for different classes, including lumen in the colon, polyps and tools. 
In~\cite{lazo2020lumen} a U-Net-like architecture based on residual blocks for lumen segmentation in ureteroscopy images is proposed. 
However, these DL-based approaches in the field of CAI only use single frames, which dismisses the chance of obtaining extra information from temporal features.

\newcommand{\widthfigone}{2.1cm}
\newcommand{\heighfigone}{2.1cm}

The exploitation of spatial-temporal information has shown to
obtain better performances than approaches that only process single frames. In~\cite{colleoni2019deep} a model based on 3D convolutions is proposed for the task of tool detection and articulation estimation, and in~\cite{moccia2019preterm} a method for infants limb-pose estimation in intensive care uses 3D Convolutions to encode the connectivity in the temporal direction. 

Additionally, recent results in different biomedical image segmentation challenges have shown the effectiveness of DL ensemble models, such as and in~\cite{wang2020automatic} where an ensemble consisting of 4 UNet-like models and one Deeplabv3+ network was proposed obtaining the 2nd place in the 2019 SIIM-ACR pneumo-thorax challenge, and in in~\cite{zheng2019new} where an ensemble which analyzed single-slices data 3D volumetric data separately was presented, obtaining top performance in the HVSMR 3D Cardiovascular MRI in Congenital Heart Disease 2016 challenge dataset.
\begin{figure}[tbp]
    \centering
    \begin{subfigure}[b]{0.99\textwidth}
            \includegraphics[width=0.99\textwidth]{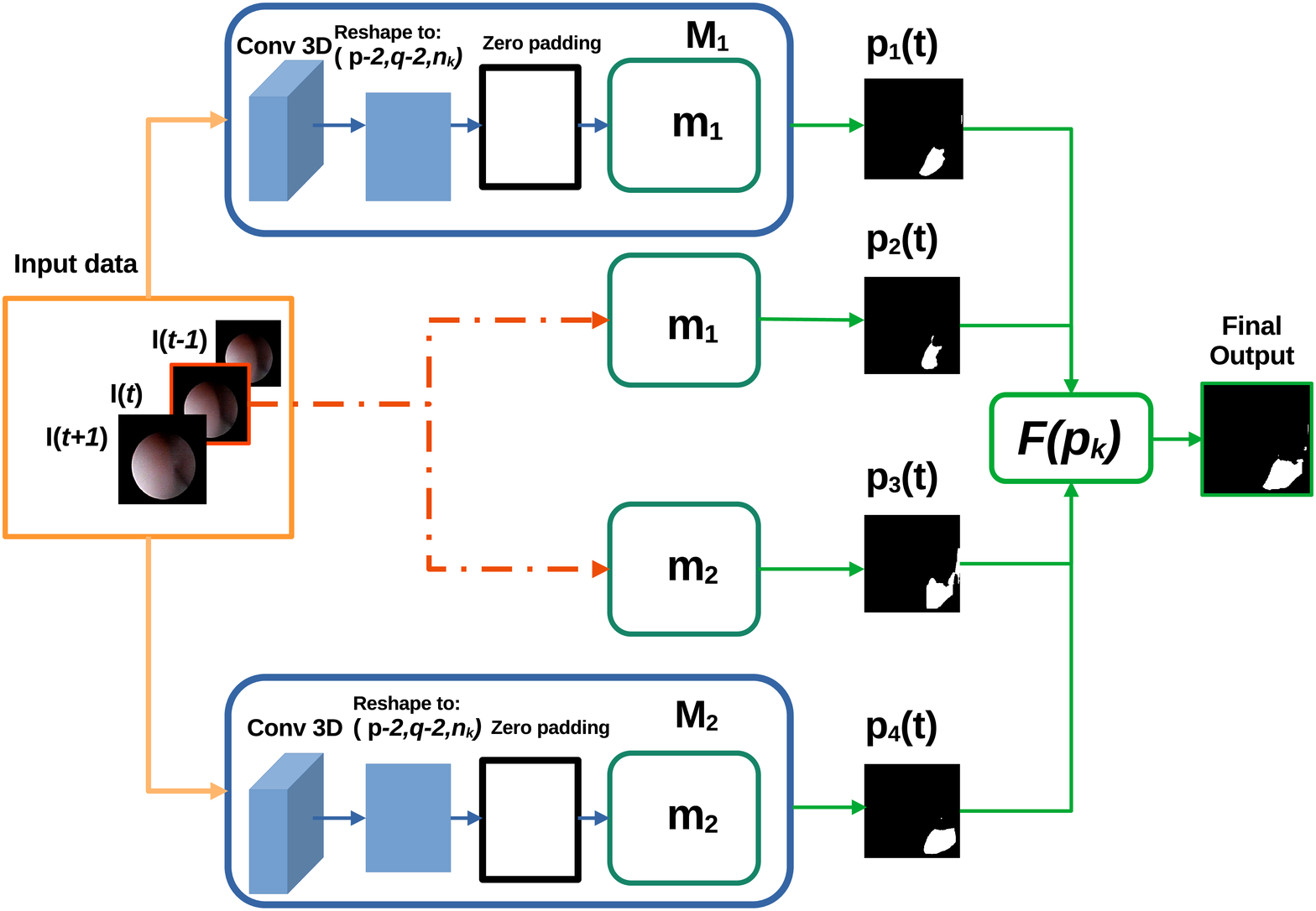}
            \footnotesize{
            \caption{The general workflow. 
            Blocks of 3 consecutive frames $I(t-1), I(t), I(t+1)$ of size $p \times q \times n_c$ (where $p$ and $q$ refers to the spatial dimensions and $n_c$ to the number of channels of each individual frame) are fed into the ensemble. Models \textbf{$M_1$} and \textbf{$M_2$} (orange line) take directly this blocks as input whereas models \textbf{$m_1$}  and \textbf{$m_2$} only take the central frame (red line). 
            Each of the $p_i(t)$ predictions made by each model are ensemble with the function $F(p_k)$ defined in Eq.~\ref{eq:ensemble} to perform the final output.  
            }}
    \end{subfigure}
    \begin{subfigure}[b]{0.49\textwidth}            \includegraphics[width=0.99\textwidth]{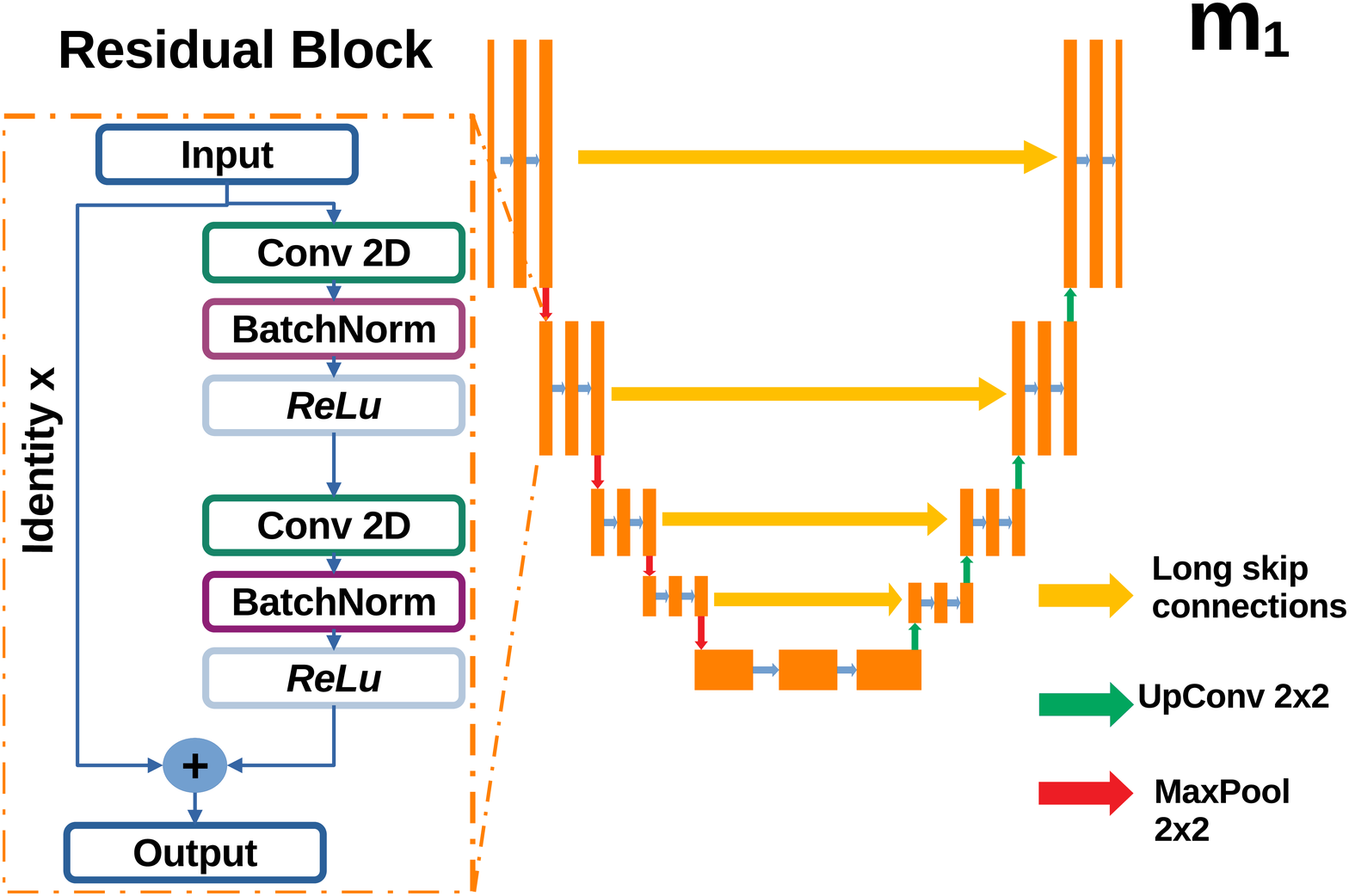}
            \caption{\textbf{$m_1$}: U-Net based on residual blocks.}
    \end{subfigure}
    \begin{subfigure}[b]{0.49\textwidth}            \includegraphics[width=0.99\textwidth]{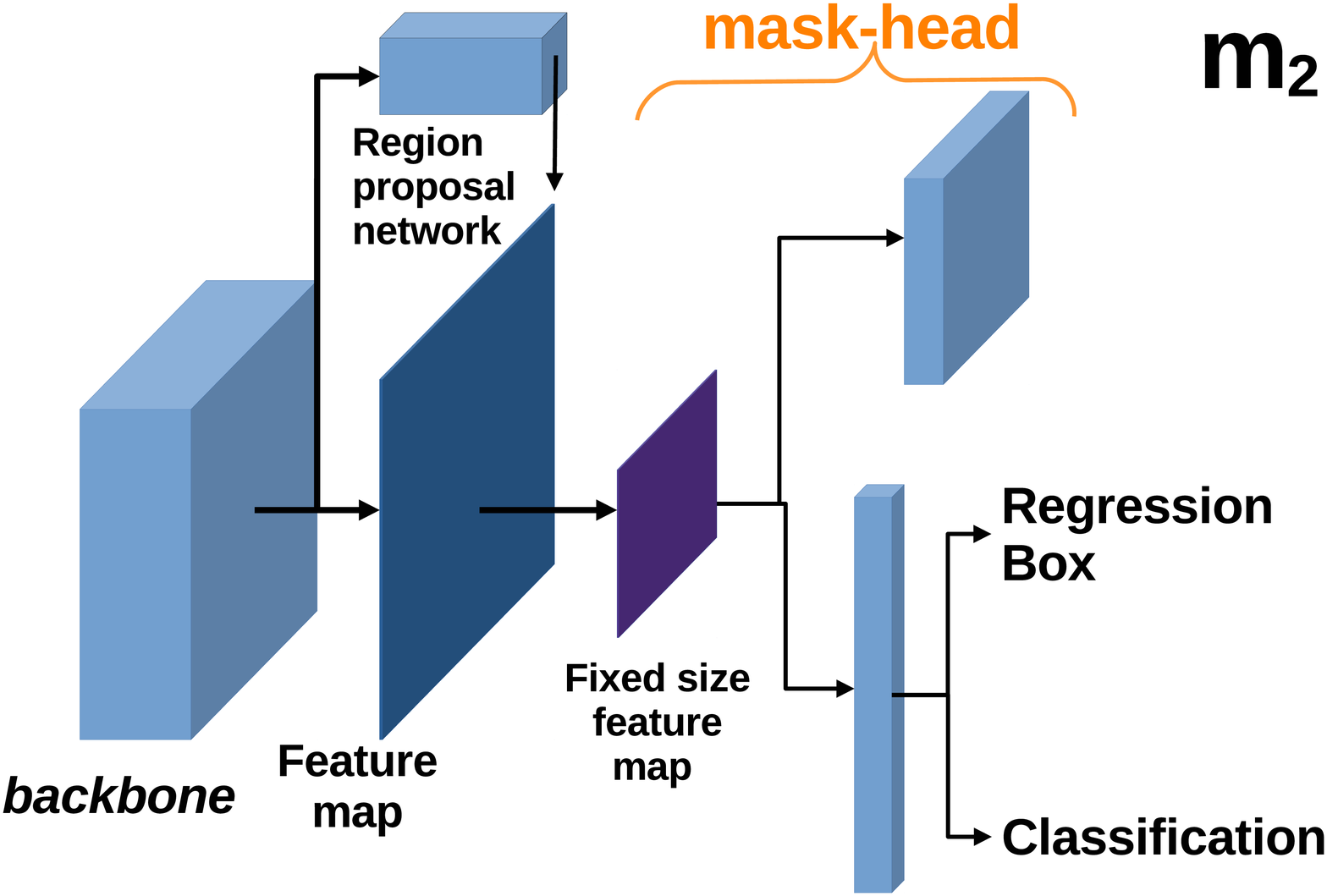}
            \caption{\textbf{$m_2$}: Mask-RCNN}
    \end{subfigure}
    \caption{
    \footnotesize{Diagram of the proposed models and their constitutive parts. The two core models $m_1$ and $m_2$ are U-Net based in residual blocks (Fig.~\ref{fig:model_1}(b)) and Mask-RCNN (Fig.~\ref{fig:model_1}(c)) respectively. In the case of U-Net based with residual blocks the dashed square depicts the composition of the residual block used. The right branch is composed of two consecutive sets of 2D Convolution layers, with its respective Batch Normalization layer and \emph{ReLu} as activation function. The output of the block is defined by the addition of the identity branch and the former branch.
    }}
    \label{fig:model_1}
\end{figure}

Inspired by both paradigms our research hypothesis is that the 
use of ensembles which use both, single-frame and consecutive-frames information could achieve a better generalization in data than models which uses only one of them. 
For this purpose we propose an ensemble model which uses in parallel 4 Convolutional Neural Networks which can exploit the information contained in single-frame and continue-frames, of ureteroscopy videos.    

\section{Proposed Method}
\label{sec:method}

As introduced in~\cite{vuola2019mask,wang2020automatic}, we considered the use of ensembles to reach a better generalization of the model when testing it on unseen data.
The proposed ensemble of CNNs for ureter's lumen segmentation is depicted in Fig.~\ref{fig:model_1}. 
Our ensemble is fed with three consecutive frames $\left[I(t-1), I(t), I(t+1)\right]$  and produces the segmentation for the frame $I_t$. 
The ensemble is made of two pairs of branches. One pair (the red one in Fig.~\ref{fig:model_1}) consists of U-Net with residual blocks ($m_1$) and Mask-RCNN ($m_2$), which process the central frame $I_t$. The other pair (orange path in Fig.~\ref{fig:model_1}) processes the three frames with $M_1$ and $M_2$, which extend $m_1$ and $m_2$ as explained in Sec.~\ref{subsec:proposed_temp}. 

It is important to notice that frames constituting the input for any $M$ are expected to have the minimal possible changes, but still significant to provide extra information which could not be obtained by other means. 
Some specific examples in our case study include the appearance of debris crossing rapidly the FOV, the sudden appearance or  disappearance of some image specularity, a slightly change in the illumination or the position of the element we are interested to segment. For this reason, we consider only three consecutive frame $I_{t-1}, I_{t}, I_{t+1}$ as input for the model. 

The core models $m_1$, $m_2$ on which our method is based are two state of the art architectures for instance segmentation: 

\begin{enumerate}
    \item ($m_1$): The \emph{U-Net} implementation used in this work is  based on residual units as used in~\cite{lazo2020lumen}, instead of using the classical convolutional blocks, this is meant to to address the degradation as proposed in~\cite{he2016deep}. 

    \item ($m_2$): Is an implementation of \emph{Mask-RCNN}~\cite{he2017mask} using ResNet50 as backbone. Mask-RCNN is composed of different stages. The fist stage is composed of two networks: a ``backbone'', which performs the initial classification of the input given a pretrained network, and a region proposal network. 
    The second stage of the model consists of different modules which include a network that predicts the bounding boxes, an object classification network and a FCN which generate the masks for each RoI.
\end{enumerate}

Since our implementation is made of different sets of models, the final output is determined using an ensemble function $F\left(p_i(t)\right)$ defined as:
\begin{equation}
    F\left(p_i(t)\right) = \frac{1}{k} \sum^{k}_i p_i(t)
\label{eq:ensemble}
\end{equation}
where $p_i(t)$ corresponds to the prediction of each of the $k=4$ models for a frame $I(t)$. 
\begin{figure}[tbp]
    \centering
    \includegraphics[width=0.8\textwidth]{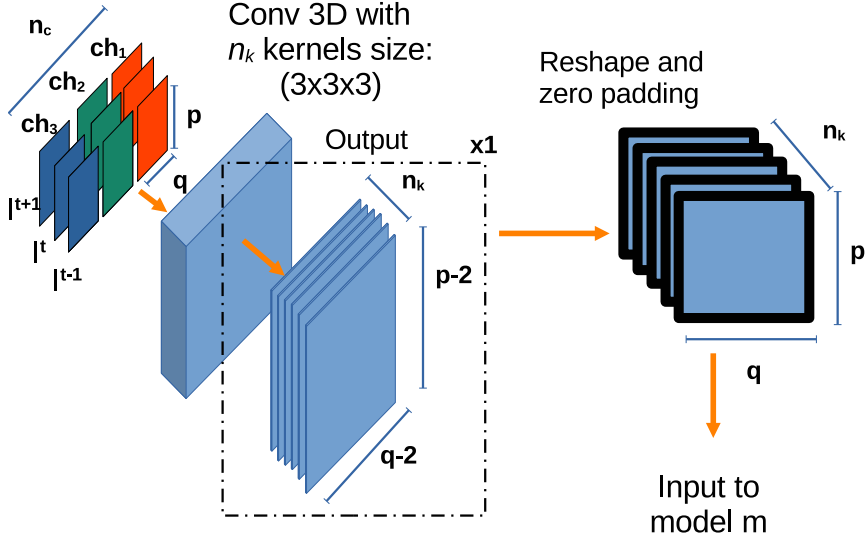}
            \caption{
            \footnotesize{The initial stage of the models \textbf{M}. 
            The blocks of consecutive frames $I(t-1), I(t), I(t+1)$ of size $p \times q \times n_c$ (where $p$ and $q$ refers to the spatial dimensions and $n_c$ to the number of channels ($ch$) of each individual frame) pass through an initial 3D Convolution with $n_k$ number of kernels. The output of this step has a shape of size $(1, p-2, q-2, n_k)$ which is padding with zeros in the 2nd and 3rd dimensions to latter, and then reshaped to fit as input for the \textbf{m} core-models}}
    \label{fig:conv3d_explain}
\end{figure}

\subsection{Extending the core models for handling multi-frame information}
\label{subsec:proposed_temp}
For each core model $m$, an extension $M$ is obtained by adapting the architecture for processing multi-frame information.

Let $\mathcal{I}$ be an ordered set of $n$ elements $I \in \mathbb{N}^{p,q,n_c}$ corresponding to frames of a video, where $p$ and $q$ represent spatial dimensions and $n_c$ the number of color channels (Fig.~\ref{fig:conv3d_explain}).
Starting from any core model ($m$), which takes as input elements from $\mathcal{I}$, we can define another segmentation model ($M$) which receives multi-frame information from $\mathcal{I}$. Specifically, it receives inputs of the form $I \in \mathbb{N}^{r,p,q,n_c}$, where $r=3$ represent the temporal dimension (number of frames). To this aim, the core model $m$ is extended by prepending an additional 3D Convolution layer with $n_k$ kernels of size ($r\times3\times3$). The new layer produces an output $H \in \mathbb{N}^{1, p-2, q-2, n_k}$, so that feeding it into $m$ is straightforward. The issue of having $p-2$ and $q-2$ instead of $p$ and $q$ after the 3D Convolution is fixed by padding the output with zeros in the two spatial dimensions. A graphical representation of the process is shown in Fig.~\ref{fig:conv3d_explain}.

\section{Evaluation}
\label{subsec:eval}

\subsection{Dataset}
\label{subsec:dataset}

For this study, 11 videos from 6 patients undergoing ureteroscopy procedures were collected. 
Videos from five patients were used for training the model and tuning hyperparameters. 
Videos from the remaining patient, randomly chosen, were kept aside and only used for evaluating the performance.
The videos were acquired from the European Institute of Oncology (IEO) at Milan, Italy following the ethical protocol approved by the IEO and in accordance with the Helsinky Declaration. 

%
The number of frames extracted and manually segmented by video is shown in Table~\ref{tab:dataset}. 
Data augmentation was implemented before starting the trainings. The operations used for this purpose were rotations in intervals of 90$^{\circ}$, horizontal and vertical flipping and zooming in and out in a range of $\pm$~2$\%$ the size of the original image. 

\newcolumntype{C}[1]{>{\centering\arraybackslash}p{#1}}
\begin{table}[tbp]
        \centering
        \caption{Information about the dataset collected. The video marked in bold indicates the patient-case that was used for testing.}
        \begin{tabular}{c c C{1.5cm} C{1.5cm} }
            Patient No. & Video No. & No. of annotated frames  & Image Size (pixels) \\ \hline \hline
             1 & Video 1  & 21   & 356x256  \\ 
             1 & Video 2  & 240  & 256x266  \\ 
             2 & Video 3  & 462 &  296x277  \\ 
             2 & Video 4  & 234 &  296x277  \\ 
             3 & Video 5  & 51 &  296x277   \\  
             4 & Video 6  & 201 &  296x277  \\ 
            \textbf{5} & \textbf{Video 7}  & \textbf{366} & \textbf{256x262}  \\ 
             6 & Video 8  & 387 & 256x262   \\ 
             6 & Video 9  & 234 & 256x262   \\ 
             6 & Video 10 & 117 & 256x262   \\ 
             6 & Video 11 & 360 & 256x262   \\ \hline
             Total  & - & 2,673  & -  \\ 
        \end{tabular}
        \label{tab:dataset}
    \end{table}

\subsection{Training Setting}
\label{subsec:training_setting}
All the models were trained, once at time, at minimizing the loss function based on the Dice Similarity Coefficient ($L_{DSC}$) defined as: 
\begin{equation}
    L_{DSC} = 1- \frac{2 TP}{2TP + FN + FP}
    \label{eq:dice_loss}
\end{equation}
where $TP$ (True Positives) is the number of pixels that belong to the lumen, which are correctly segmented, $FP$ (False Positives) is the number of pixels miss-classified as lumen, and $FN$ (False Negatives) is the number of pixels which are classified as part of lumen but actually they are not. 

For the case of ($m1$) the hyperparameters learning rate (lr) and mini batch size (bs) were determined using a 5-fold cross validation strategy with the data from patients 1, 2, 3, 4 and 6 in a grid search. 
The ranges in which this search was performed were $lr = \lbrace 1e-3, 1e-4, 1e-5, 1e-6 \rbrace$  and $bs = \lbrace 4,8,16 \rbrace$. 
The $DSC$ was set as the evaluation metric to determine the best model for each of the experiments. 
Concerning the extensions $M$, the same strategy was used to determine the number of kernels of the input 3D convolutional layer. The remaining hyperparameters were set the same as for $m_1$.

In case of $m_2$, the same 5-fold cross validation strategy was used. The hyperparameters tuned were: the backbone (from the options ResNet50 and ResNet101~\cite{he2016deep}) and the value of minimal detection confidence in a range of 0.5 to 0.9 with differences of 0.1. 
To cover the range of different sizes of masks in the training and validation dataset the anchor scales were set to the values of 32, 64, 128 and 160. 
In this case the number of filters in the initial 3D convolutional layer was set to a value of 3 which is the only one that could match the predefined input-size, after reshaping, of ResNet backbone. 

For each core models and their respective extensions, once the hyperparameters values were chosen, an additional training process was carried out using these values in order to obtain the final model. The training was performed using all the annotated frames obtained from the previously mentioned 5 patients, 60$\%$ of the frames were used for training and 40$\%$ for validation.
The results obtained in this step were the ones used to calculate the ensemble results the function defined in Eq.~\ref{eq:ensemble}.

The Networks were implemented using \textit{Tensorflow} and \textit{Keras} frameworks in Python 3.6 trained on a \textit{NVIDIA GeForce RTX 280} GPU.

\subsection{Performance Metrics}
\label{subsec:performance_metrics}

The performance metrics chosen were $DSC$, Precision ($Prec$) and Recall ($Rec$), defined as:
\begin{equation}
    DSC = 1 - L_{DSC}
    \label{eq:DSC}
\end{equation}
\begin{equation}
    Prec = \frac{TP}{TP+FP}
    \label{eq:Prec}
\end{equation}
\begin{equation}
    Rec = \frac{TP}{TP+FN}
    \label{eq:Rec}
\end{equation}

\subsection{Ablation study and comparison with sate-of-the-art}
\label{subsec:ablation}

First, the performance of the proposed method was compared with the one presented in~\cite{lazo2020lumen}, where the same U-Net based on residual blocks architecture was used. Then, as ablation study, four versions of the ensemble model were tested:
\begin{enumerate}
    \item ($m_1$,$m_2$): only single-frame information was considered in the ensemble;
    \item ($M_1$,$M_2$): only multi-frame information was considered in the ensemble;
    \item ($m_1$,$M_1$), ($m_2$,$M_2$): each of the core models, and its respective extension, were considered in the ensemble, separately.
\end{enumerate}
In these cases, the ensemble function was computed using the values of the predictions of each of the models. The Kruskal-Wallis test on the $DSC$ was used to determine the statistical significance between the different single models tested. 
\begin{figure}[tbp]
    \centering
    \includegraphics[width=0.85\textwidth]{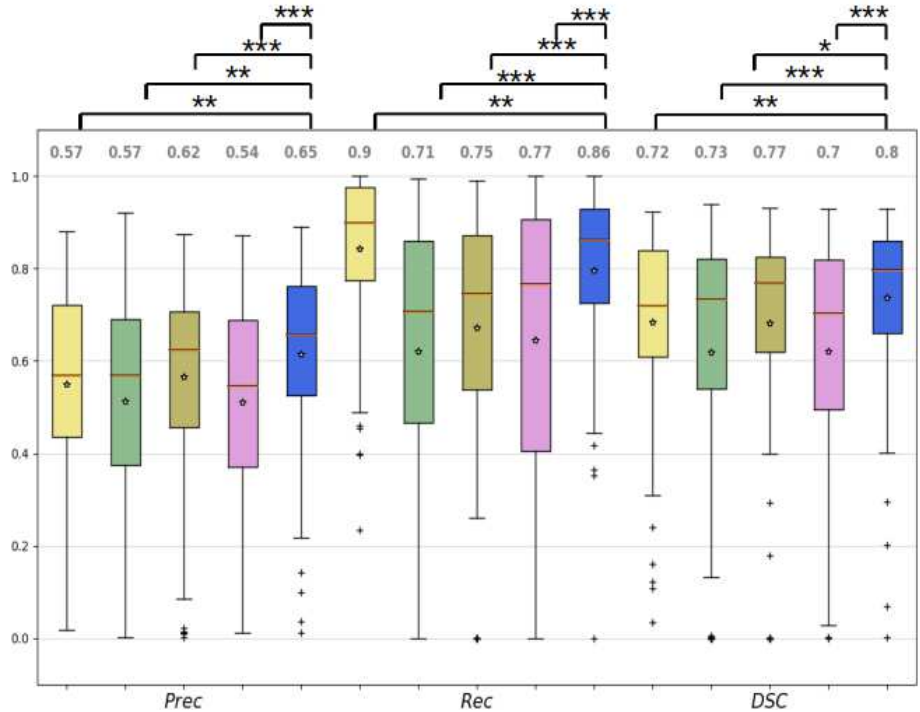}
    \caption{
    \footnotesize{
    Box plots of the precision ($Prec$), recall ($Rec$) and the Dice Similarity Coefficient ($DSC$) for the models tested.  
    $m_1$~(yellow): ResUNet with single image frames, 
    $m_2$~(green): ResUNet using consecutive temporal frames, $M_1$~(brown): Mask-RCNN with single image frames, 
    $M_2$~(pink): Mask-RCNN using consecutive temporal frames, and the proposed ensemble method (blue) formed by all the previous models. 
    The asterisks represent the significant difference between the different architectures in terms of the Kruskal-Wallis sign rank test (* $p<0.05$, ** $p<0.01$, *** $p<0.001$).} }
    \label{fig:results_dsc}
\end{figure}

\section{Results}
\label{sec:results}
\begin{figure}[tbp]
    \centering
    \includegraphics[width=0.81\textwidth]{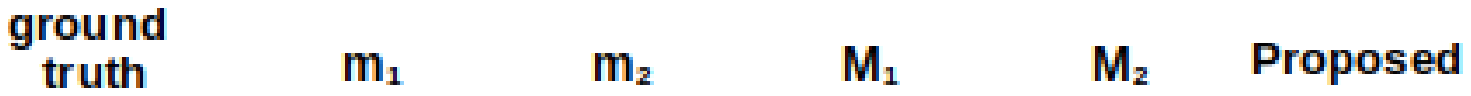}
    \includegraphics[width=0.81\textwidth]{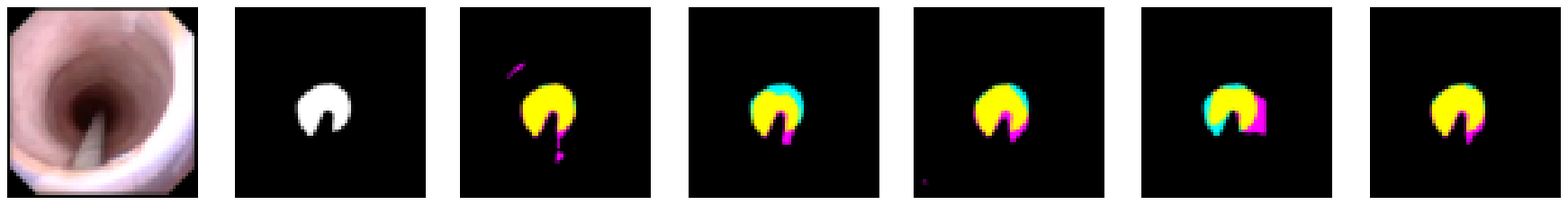}
    \includegraphics[width=0.81\textwidth]{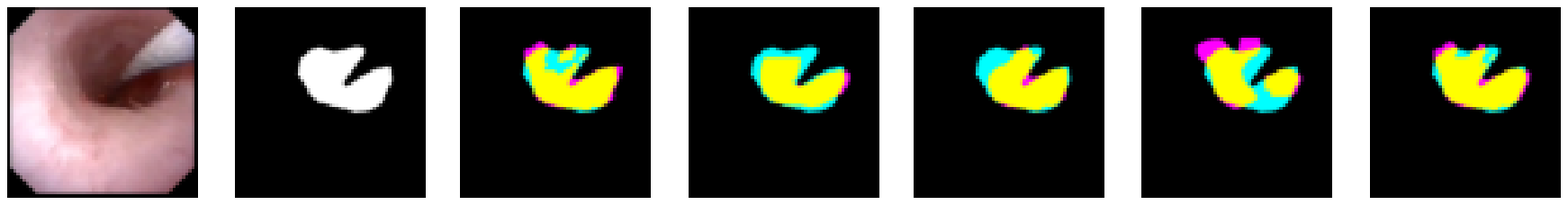}
    \includegraphics[width=0.81\textwidth]{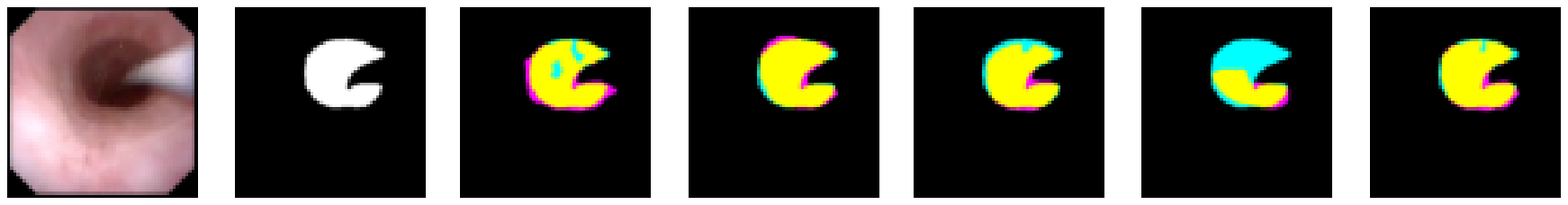}
    \includegraphics[width=0.81\textwidth]{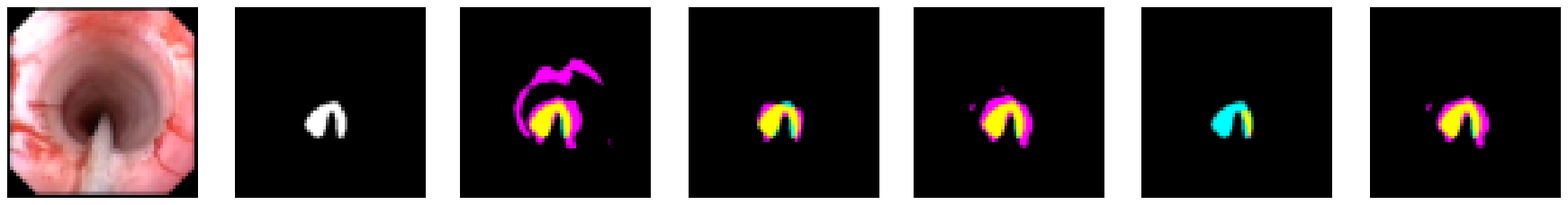}
    \includegraphics[width=0.81\textwidth]{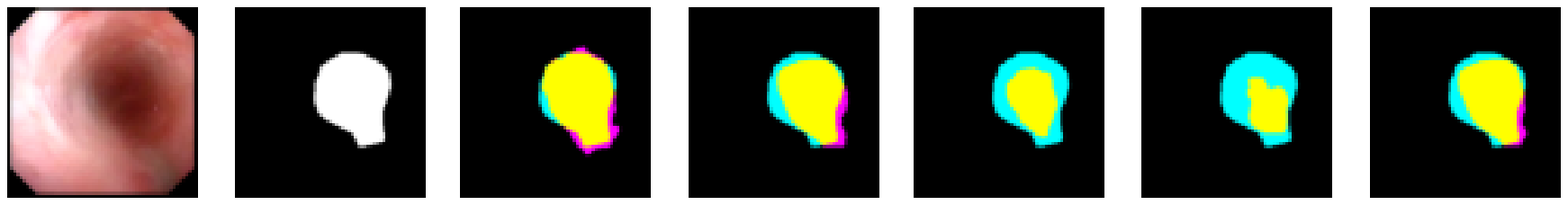}
    \includegraphics[width=0.81\textwidth]{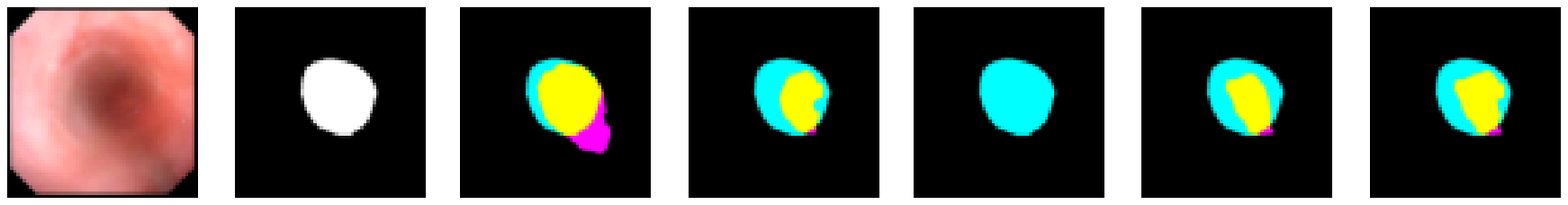}
    \includegraphics[width=0.81\textwidth]{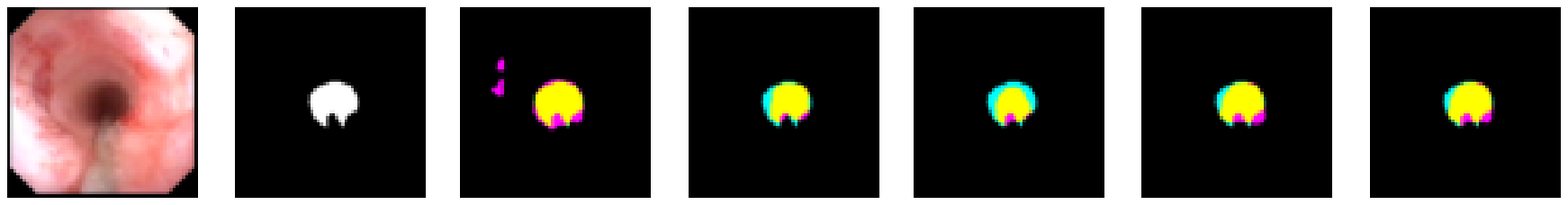}
    \includegraphics[width=0.81\textwidth]{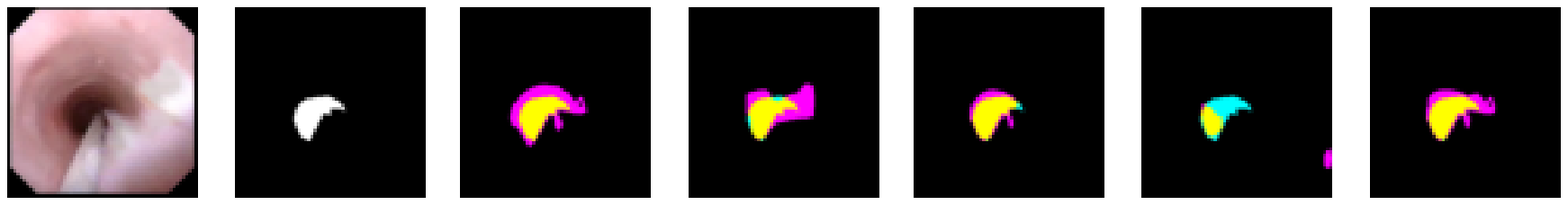}
    \includegraphics[width=0.81\textwidth]{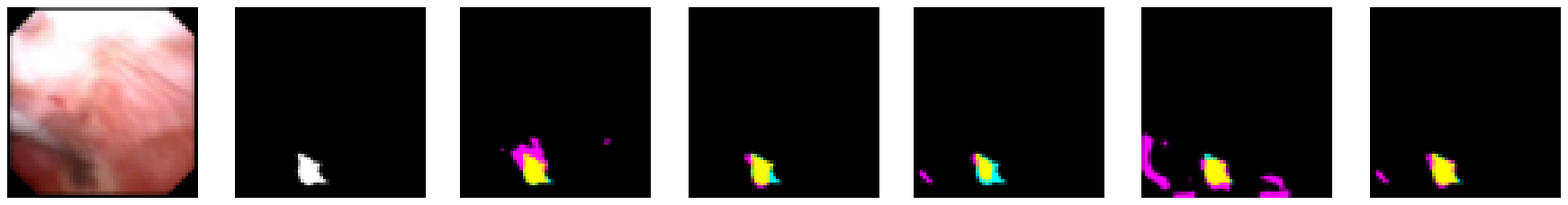}
    \includegraphics[width=0.81\textwidth]{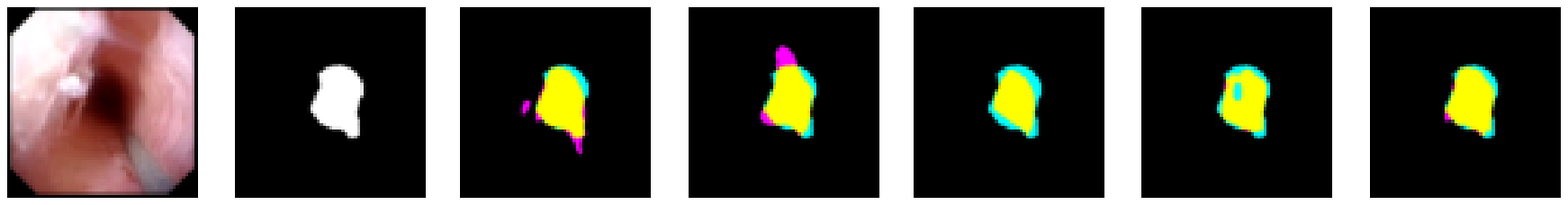}
    \includegraphics[width=0.81\textwidth]{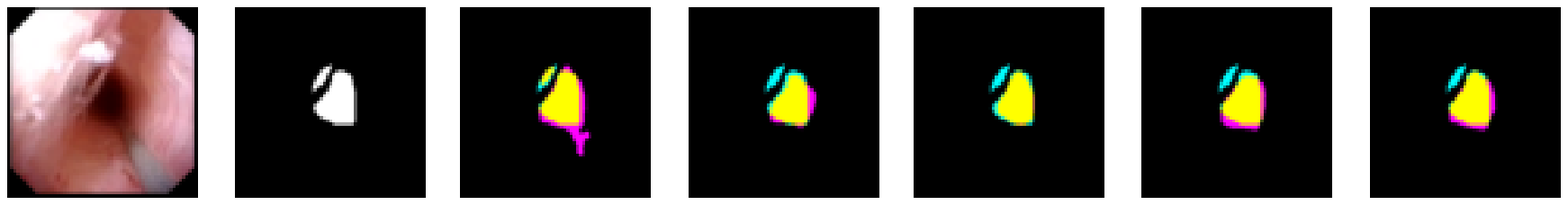}
    \caption{
    \footnotesize{
    Samples of segmentation with the different models test. 
    The colors in the Overlay images represent the following for each pixel. 
    True Positives~(TP): Yellow, False Positives~(FP): Pink, False Negatives~(FN): Blue, True Negatives~(TN): Black. 
    The first three rows depict images where the lumen is clear with the respective segmentation from each model. Rows 4-7 show cases in which some kind of occlusion appears. Finally the rows 8-12 depict cases in which the lumen is contracted, and/or there is debris crossing the FOV. 
    }
    }
    \label{fig:samples_results_segmentation}
\end{figure}

\begin{table}[tbp]
    \centering
    \caption{
    \footnotesize{Average Dice Similarity Coefficient ($DSC$), precision ($Prec$) and recall ($Rec$) in the cases in which the ensemble were formed only by: 1. Spatial models $(m_1, m_2)$; 2. spatial-temporal $(M_1, M_2)$, 3. ResUnet with both spatial and temporal inputs $(M_1, m_1)$ and 4. Mask-RCNN with the same setup $(M_2, m_2)$. $F(*)$ refers to the ensemble function used Eq.~\ref{eq:ensemble}, and the components used to form the ensemble are stated between the parenthesis.}}
    \begin{tabular}{c|c|c|c}
    F(*)     & $DSC$ & $Prec$ & $Rec$ \\ \hline \hline
    ($m_1$, $m_2$) & 0.78 & 0.65 & 0.71\\ 
    ($M_1$, $M_2$) & 0.71 & 0.55 & 0.57\\ 
    ($M_1$, $m_1$) & 0.72 & 0.56 & 0.66\\ 
    ($M_2$, $m_2$) & 0.68 & 0.51 & 0.63\\ 
    \end{tabular}
    \label{tab:results_ablation}
\end{table}
The box plots of the $Prec$, $Rec$ and the $DSC$ are shown in Fig.~\ref{fig:results_dsc}. Results of the ablation study  are shown in Table~\ref{tab:results_ablation}.
%
The proposed method achieved a $DSC$ value of 0.80 which is $8\%$ better than $m_1$ using single frames ($p<0.01$) and $3\%$ than $m_2$ trained as well with single frames ($p<0.05$). 
%
When using single-frame information, $m_2$ performs $5\%$ better than $m_1$. However the results is the opposite using multi-frame information.
%
The ensembles of single-frame models ($m_1,m_2$) performs $7\%$ better with respect to ensembles of models exploiting multi-frame information ($M_1,M_2$).
%
In the case of spatio-temporal-based models U-Net based on residual blocks ($M_1$) performs $3\%$ better than the one based on Mask-RCNN ($M_2$).    
This might be due to the constraint of fitting the output of the 3D Convolution into the layers of the backbone of Mask-RCNN. 
%
The same limitation might explain the similar behaviour when it comes to the comparison of the ensembles composed only by U-Net based in residual blocks models and Mask-RCNN-based models, where the former one performs $4\%$ better than the second one. 
The only model which achieves a better performance than the proposed one in any metric is U-Net based on residual blocks with the $Rec$, obtaining a value 0.04 better than the model we proposed.
Visual examples of the achieved results are shown in Fig.~\ref{fig:samples_results_segmentation} and in the video attached to this paper. 
Here, the first 2 rows show frames in which the lumen appears clearly and there is no presence of major image artifacts. 
As observable, each single model underestimate the ground-truth mask. 
However, their ensemble gives a better approximation. 
The next 2 rows show cases in which some kind of occlusions (such as blood or debris) is covering most of the FOV. In those cases, single-frame models ($m$) give better results than its counterparts handling temporal information ($M$). 
Finally, the last 2 rows of the image contain samples showing minor occlusions (such as small pieces of debris crossing the FOV) and images where the lumen is not on focus. 

The average inference time was also calculated. Results for $m_1$ and $M_1$ are 26.3$\pm$3.7~ms and 31.5$\pm$4.7~ms, respectively. 
In case of $m_2$ and $M_2$, the average inference times are 29.7$\pm$2.1~ms and 34.7$\pm$6.2~ms, respectively.
In the case of the ensemble, the average inference time was 129.6$\pm$6.7~ms when running the models consecutively.

\section{Discussion}
\label{sec:disc}
The proposed method achieved satisfactory results, outperforming existing approaches for lumen segmentation~\cite{lazo2020lumen}. Quantitative evaluation, together with a visual inspection of the obtained segmentations, highlight the advantage of using ensembles, confirming our research hypotheses.  
%
%
This is particularly appreciable in presence of occlusions such as blood or dust covering the FOV (Fig.~\ref{fig:samples_results_segmentation} rows 4-7). In those cases, single-frame-based models tended to include non-lumen regions in the predicted segmentation. An opposite behavior was observed when using only multi-frame-based models, which tended to predict smaller regions with respect to the ground-truth and which is also noticeable in the general performances carried during the ablation studies (Table~\ref{tab:results_ablation}).
The ensemble of all of them resulted, instead, in a predicted mask closer to the ground-truth and exemplifies why the use of it in general turns into better performances.
It was also observed that the proposed ensemble method was able to correctly manage undesirable false positives appearing in single models. This is due the fact that those false positives did not appear in all the models at the same regions, therefore, the use of ensembles eliminate them from the final result.  
This is of great importance in the clinical practice, given that false positive classifications during endoluminal inspection might results in a range of complications of the surgical operation, including tools colliding with tissues~\cite{he2020endoscopic}, incorrect path planning~\cite{alsunaydih2020navigation}, among others.

Despite the positive results achieved by the proposed approach, some limitations are worth to be mentioned. 
Computational time required for inference is one of those. 
In terms of inference time, the proposed model requires 4 times more than previous implementations.
However, it is important to state that when it comes to applications of minimal invasive surgery, accuracy may be preferred over speed
to avoid any complication, such as perforations of the ureter~\cite{delaRosette_2006}.
Furthermore, such time could be improved by taking advantage of distributed parallel set-ups. 
%
A final issue is related to the scarcity of public available and annotated data, necessary to train and benchmark, which is a well-known problem in literature.
However, this can be overcome in future as new public repositories containing spatial-temporal data are released. 


\section{Conclusion}
\label{sec:conclusion}
In this paper, we introduced a novel ensemble method for ureter's lumen segmentation. 
Two core models based on U-Net and Mask-RCNN were exploited and extended, in order to capture both single-frame and multi-frame information. 
Experiments showed that the proposed ensemble method outperforms previous approaches for the same tasks~\cite{lazo2020lumen}, by achieving an increment of $7\%$ in terms of $DSC$. 
In the future, evident extensions of the present work will be investigated, including better methods to fit spatial-temporal data into models which were pre-trained in single image datasets (such as Mask-RCNN).
Furthermore, we will investigate methods for decreasing the inference time, thus allowing real-time applications.

%

\bibliographystyle{ieeetr}      
\bibliography{references}

\end{document}